\documentclass[11pt]{article}
\usepackage[a4paper,hmarginratio=1:1,vmarginratio=2:3,totalwidth=16cm,
  totalheight=20.8cm]{geometry}  
\usepackage{latexsym}

\usepackage{epsfig}

\newcommand{\cJ}{{\cal J}}

\newcommand{\cL}{{\cal L}}

\newcommand{\cO}{{\cal O}}

\newcommand{\bZ}{{\bf Z}}

\newcommand{\ra}{\rightarrow}
\newcommand{\be}{\begin{equation}}
\newcommand{\ee}{\end{equation}}
\newcommand{\bea}{\begin{eqnarray}}
\newcommand{\eea}{\end{eqnarray}}

\DeclareMathSymbol{\mg}{\mathrel}{symbols}{"1D}

\newcounter{oldcounter} 
\addtocounter{equation}{1}
\begin{document} 
\begin{flushright}  
{DAMTP-2005-70}\\
{DESY-2005-133}
\end{flushright}  
\vskip 1.7cm
\begin{center} 
{\Large 
{\bf  Higher Derivative Operators as Counterterms\\}
\vspace{0.3cm}
{\bf in Orbifold  Compactifications.\\} }
\vspace{1.cm} 
{\bf D.M. Ghilencea $^a$} and
{\bf  Hyun Min Lee $^b$}
\\
\vspace{0.8cm} 
$^a$ {\it D.A.M.T.P., Centre for Mathematical Sciences, 
University of Cambridge, \\
Wilberforce Road, Cambridge CB3 OWA, United Kingdom}\\
\vspace{0.4cm}
$^b$ {\it Deutsches Elektronen-Synchrotron DESY, \\
D-22603 Hamburg, Germany}\\
\end{center}
\begin{abstract}
\noindent
In the context of 5D N=1 supersymmetric models compactified
on $S_1/Z_2$ or $S_1/(Z_2\times Z_2')$ orbifolds and with
brane-localised superpotential, higher derivative operators are
generated  radiatively as  one-loop counterterms to the mass of
the (brane or zero mode of the bulk) scalar field. It is shown that 
the presence of such operators which are brane-localised  is not 
related to the mechanism of supersymmetry breaking considered (F-term, 
discrete or continuous Scherk-Schwarz breaking) and initial
supersymmetry does not protect against the dynamical generation of 
such operators. Since in many realistic models the scalar field is 
commonly regarded as the Higgs field, and the higher derivative 
operators seem a generic presence in orbifold compactifications, 
we stress the importance of these operators for solving the hierarchy 
problem.
\vfill{
\noindent
{\it Contribution to the\\
 Bad Honnef meeting, ``Beyond the
  Standard Model'', March 2005,  Bad Honnef, Germany,  
 ``Planck 2005'', May 2005,  Trieste, Italy and
 ``Supersymmetry 2005'', July 2005, Durham.}}
\end{abstract}

\newpage

\section{Introduction}

In recent years the study of radiative corrections from 
compact dimensions has seen a strong development in the context 
of field theory approaches to orbifold compactifications.
The interest in  such studies has theoretical
and  experimental motivations. First, one cannot exclude 
the possibility that radiative
corrections from compact dimensions may have some experimental
signatures in the context of  the  Large Hadron Collider experiment. 
  Another motivation is  that such 
studies of compactification allow a comparison with the more 
comprehensive approaches  of  string theory. 
Field theory orbifolds can give one-loop results similar to those of the
string  in the limit $\alpha'\ra 0$ \footnote{for an  example see 
\cite{Ghilencea:2002ff}, \cite{Ghilencea:2003kt}.}. This  provides a better
understanding of compactification and an indication whether string theory  
can provide an UV completion for effective field theory models.
Finally, field theory orbifolds allow us to re-address in a consistent
framework  well known problems such as the hierarchy problem,
supersymmetry or electroweak symmetry breaking, etc, whether or not
there is a link with string theory.

An aspect that is often overlooked in the studies of radiative
corrections induced by compactification (to the couplings or the
masses in the theory) is the role of the higher  (dimension) 
derivative operators\footnote{For some
  studies of higher derivative operators see for example 
 \cite{Pais:1950za}, \cite{Kazakov:2002jd}.}.
 This observation applies to both field theory 
and string theory orbifolds. The study of such operators is
particularly  compelling when they are generated as counterterms in
the action, to ensure the quantum consistency of the model under
study. Specific examples where higher derivative operators are a
result of the compactification can be found in 
\cite{Oliver:2003cy}-\cite{dmghml}.
In general one could  assume that  the effects of such operators 
are suppressed at low energies.  From a 4D 
perspective this suppression is ensured by the compactification scale 
which is the   natural 4D cutoff scale.  If this scale is low,  such 
operators are less suppressed and  affect the radiative 
corrections.  Their effects become manifest as we increase the
external momentum $q^2$ of the  Green function under 
study,  relative to the compactification scale $1/R$, from $q^2\ll
1/R^2$ to regions where $q^2\gg 1/R^2$. Here we  
assume that  $1/R$ and $q^2$ have arbitrary (fixed) values, to
allow a general study of the effects of these operators,  whether 
$q^2  R^2\!\ll\! 1$ or $q^2  R^2\!\gg\! 1$.

The importance of the study of 
 higher derivative operators in gauge theories on
orbifolds is due  to the underlying physics attached to them. 
In the absence of 
a UV completion of the effective field theory which 
provides the framework of our study, such operators can be present
with unknown coefficients, thus affecting the predictive power of the models. 
Further complications can arise, such as the presence of ghosts
 fields,  unitarity  violation,
non-locality effects, etc, \cite{Pais:1950za,Kazakov:2002jd}
 and  for these reasons, in general theories with higher
 derivative operators were  not extremely  popular.
 It is thus even more important to address the consequences of the 
presence of such operators in gauge theories on orbifolds, where such 
operators are generated radiatively by gauge
\cite{Oliver:2003cy,Ghilencea:2003xj,GrootNibbelink:2005vi,dmghml} or
 Yukawa  interactions \cite{Ghilencea:2005hm,Ghilencea:2004sq,dmghml}. 
Their investigation  has many implications for model building.

One would expect to find out more about the role of higher derivative
operators from string calculations. Unfortunately  this is not always the
case, and sometimes little light is shed
on the corrections such operators 
induce, partly due to the on-shell formulation of the string.
More explicitly, in the context of string loop corrections to the
gauge couplings \cite{DKL} higher derivative counterterms can be missed
by the string approach, although they can be shown to be present 
in effective field theories 
\cite{Oliver:2003cy,Ghilencea:2003xj,GrootNibbelink:2005vi}.
This  raises questions \cite{Ghilencea:2002ak} (also
\cite{Ghilencea:2002ff}) on the 
{\it exact} matching of the two approaches to compactification:
string theory versus effective field theory.
Nevertheless,  higher derivative operators can  be studied
even in the absence of any link with string theory or of a 
compactification of a higher dimensional theory. This can be done
in the context of 4D  field theories with additional higher dimension
operators, where the scale  where they become 
relevant is regarded as the scale of new physics (rather than the
compactification scale).

Higher dimensional models of physics beyond the Standard Model (SM) 
or the Minimal Supersymmetric Standard Model   (MSSM) require in general 
some amount of supersymmetry, for reasons of stability, hierarchy problem etc.
However, such models are nevertheless  non-renormalisable, 
and then such operators can be present
in the action. It is  then interesting to address  the extent to 
which initial supersymmetry protects against the generation of such 
operators  by radiative corrections. In this context of particular
interest for the hierarchy problem is the relation between the nature 
of supersymmetry breaking on 5D orbifolds and the presence of higher 
derivative operators as  loop counterterms to the mass of a scalar 
field~\cite{Ghilencea:2005hm,Ghilencea:2004sq,dmghml}.

In the present work we address how (brane-localised)  higher
derivative operators
emerge as counterterms to the mass of the scalar field, from radiative 
corrections induced by (brane-localised) superpotentials in 5D N=1 
supersymmetric models. We review models on $S_1/Z_2$, 
$S_1/(Z_2\!\times\! Z_2')$  investigated in 
\cite{Ghilencea:2005hm,Ghilencea:2004sq} and show how higher derivative 
operators emerge from compactification regardless of the exact
details  of the mechanism for 
supersymmetry breaking.  Such interaction is  generic in the
literature, and some models for which our findings may be relevant
can be found in  refs.\cite{Antoniadis}-\cite{gnm}.
For this study we consider that after orbifolding 
the remaining N=1 supersymmetry is broken via F-term 
breaking,  discrete or continuous Scherk-Schwarz  breaking or
additional orbifolding ($Z_2'$). Our
results  show that supersymmetry does not protect  against the presence of 
higher derivative counterterms to the mass of the scalar field, 
even at the one-loop level.  The implications for the
hierarchy problem are also discussed briefly.

\section{Higher derivative operators as counterterms on orbifolds.}

The models we consider have 5D N=1 supersymmetry
and are compactified on $S_1/Z_2$ orbifolds. To a large extent  our 
considerations also apply to the $S_1/(Z_2\!\times\! Z_2')$ orbifold. 
The spectrum of the models  will contain representations of this supersymmetry.
Vector supermultiplets on $S_1/Z_2$ may be described in a 4D language 
as made of a vector superfield $V(\lambda_1,A_\mu)$ and adjoint chiral
 superfield $\Sigma((\sigma+i A_5)/\sqrt
2,\lambda_2)$ where $\lambda_{1,2}$ are Weyl fermions, $\sigma$  is a real
scalar and $A_\mu, A_5$ is the 5D gauge field. A hypermultiplet contains
$\Phi(\phi,\psi)$ and $\Phi^c(\phi^c,\psi^c)$ with opposite SM quantum
numbers, with $\phi, \phi^c$ as complex scalars and $\psi, \psi^c$
the  Weyl fermions. Under the orbifold action $y\ra -y$ 
we impose that the above fields transform as
\begin{eqnarray}\label{orbifold}
\Phi(x, - y)=\Phi(x,y),\quad  \qquad &&  V(x,-y)=V(x,y)
\nonumber\\
\qquad\Phi^c(x, -y)=-\Phi^c(x,y) \qquad && \Sigma(x,-y)=-\Sigma(x,y).
\end{eqnarray}
Here $\Phi$ is any of the SM fields $Q, U, D, L, E$ of the
Standard Model. 
As a result of (\ref{orbifold}), the initial 5D N=1 supersymmetry is broken
and  the  fixed points ($y=0, \pi R$) of the orbifold have  a
remaining  4D N=1 supersymmetry. The gauge field is 
even under the orbifold action so it has a zero mode,
which is the  massless 4D gauge boson of the model. 

In such models one would like to introduce gauge and Yukawa
 interactions. Here we restrict the discussion to the case
 of the latter, to show how higher derivative operators are
 generated at one-loop\footnote{Gauge interactions 
 generate  higher derivative operators beyond 1-loop in 5D,
or in 6D at 1-loop~\cite{Ghilencea:2003xj,GrootNibbelink:2005vi,dmghml}.}.
 Given the amount of supersymmetry in the bulk and at the fixed points,
the only option is to consider a brane-localised superpotential. The 
interaction is then

\begin{eqnarray}\label{interaction}
  \cL_4 = \int\!  dy \; \delta(y)
  \left\{-\int \!\!d^2\! \theta
 \,\, \Big[\lambda_t\, Q \,U \,H_u
      + \lambda_b \, Q \, D\,  H_d + \cdots \Big]
  + {\rm h.c.} \right\}.\\[-11pt]
\nonumber
\end{eqnarray}
The 5D coupling
$\lambda_{t}\!=\!f_{5, t}/M_*^n\!=\!(2\pi R)^n f_{4, t}$ and $f_{5,t}$
 ($f_{4, t}$) is  the dimensionless 5D  (4D) coupling,
$M_*$ is the cutoff of the theory.
In the following $Q$, $U$, $D$ superfields are assumed 
 to be bulk fields\footnote{Other possibilities for the character
   bulk/brane of the fields $Q,U$ 
are considered, see later.}, so they have mass  dimension
$[Q]\!=\![U]\!=\![D]\!=\!3/2$. We also introduced 
the Higgs fields  $H_{u,d}$. These  can be brane fields when
$[H_{u,d}]=1$ ($n=1$)  or bulk fields  $[H_{u,d}]=3/2$, ($n=3/2$) 
when they must also have  a
$H_{u,d}^c$ partner. If $H_{u,d}$ are also bulk fields they 
satisfy a condition similar to that for $\Phi$ in eq.(\ref{orbifold}).
The above spectrum and  interaction define our minimal
model\footnote{The presence of both $H_{u,d}$ is to avoid quadratic
divergences to the scalar field mass from  FI terms \cite{dsh}.}.
Such interaction is generic in 5D extensions of the SM or the MSSM 
and was extensively considered in the past  (for such 5D models see
\cite{Antoniadis}-\cite{gnm}). New effects so far
overlooked are presented below.

After the  orbifold  action (\ref{orbifold}) 
on the hypermultiplets and vector multiplets,
which breaks the 5D N=1 supersymmetry, the
remaining 4D N=1 supersymmetry  can be broken using

\vspace{0.2cm}

\noindent
{\bf 1).}  F-term supersymmetry breaking.

\noindent
{\bf 2).}  Discrete Scherk-Schwarz twists.

\noindent
{\bf 3).}  Continuous Scherk-Schwarz twists.

\noindent
{\bf 4).}  An additional orbifolding by $Z_2'$, so the orbifold is actually
  $S_1/(Z_2\times Z_2')$.

\noindent
We briefly review these cases, and then address the one-loop correction to 
the mass of $\phi_{H_u}$ induced by interaction (\ref{interaction});
(similar considerations apply for $H_d$).

\vspace{0.4cm}
\noindent
{\bf Case 1).} F-term supersymmetry breaking.
In this case one considers supersymmetry broken at a distant (hidden)
brane located at $y=\pi R$ by
\begin{eqnarray}\label{ZZ}
\cL_4=\int dy \,\delta(y-\pi R)\, \bigg\{
\int d^2\theta \,\,M_*^2 \, Z+h.c.
-\int d^4\theta \,\,\bigg[
\frac{c_Q}{M_*^3}\,Q^\dagger Q\,Z^\dagger Z+
\frac{c_U}{M_*^3}\,U^\dagger U\,Z^\dagger Z\bigg]
\bigg\},
\end{eqnarray}
where $Z$ is a (gauge singlet) brane field at $y=\pi R$ and $M_*$ is
the cutoff scale of the model.
The bulk  fields  $Q, U$ feel the   supersymmetry breaking via
couplings as in the above integral over $d^4 \theta$.
When $<Z>\sim F_Z \theta^2$ the bulk fields $\phi_{M}$, $M=Q, U$ which have
non-zero coupling at $y=\pi R$ brane have the spectrum modified, while
their fermionic partners $\psi_{Q,U}$ do not couple to $Z$ (and
neither do $\psi^c_{Q,U}$, $\phi^c_{Q,U}$, due to eq.(\ref{orbifold}))
and their Kaluza-Klein  spectrum is not affected.  The Higgs field
$H_u$ (also $H_d$)  that is considered in this case 
to be localised at $y=0$ (to avoid a direct coupling to $Z$)
feels supersymmetry  breaking at $y=\pi R$ via loops
of bulk fields $Q, U$.
As a result the Kaluza-Klein modes  $\phi_{M,k}$ have their mass shifted by
(\ref{ZZ}) \cite{Arkani-Hamed:2001mi} 
and  $m_{\psi_{M,k}}\not=m_{\phi_{M,k}}$,
for all positive $k$ including  the zero modes.

\vspace{0.4cm}
\noindent
{\bf Case 2).} Discrete Scherk-Schwarz supersymmetry breaking.
In this case 5D fields acquire under a $2 \pi R$ shift a phase which
is the R-parity charge of the fields
\begin{eqnarray}\label{rparity}
Z_{2,R} M(x,y,\theta)    =-M(x,y,-\theta),\qquad && 
Z_{2,R} M^c(x,y,\theta)  =-M^c(x,y,-\theta),\quad M=Q, U\nonumber\\
Z_{2,R} H(x,y,\theta)    =H(x,y,-\theta),\quad\qquad\! && \!
Z_{2,R} H^c(x,y,\theta)  =H^c(x,y,-\theta),\nonumber\\
Z_{2,R} V(x,y,\theta)    =V(x,y,-\theta),\qquad\quad &&\,\,\,
Z_{2,R}\Sigma(x,y,\theta)=\Sigma(x,y,-\theta)
\end{eqnarray}
In this case $H_{u,d}$ can be either  bulk or brane fields;
in the former case the second line in eq.(\ref{rparity}) applies and 
stands for
both $H_{u,d}, H^c_{u,d}$. As a result of (\ref{rparity}) 
 $\phi_{M,k}$ and $\psi_{M,k}$ ($M=Q, U$) and in particular their zero
modes  acquire different masses.  The field $\phi_{H_u}$ (also $\phi_{H_d}$) 
-~or its zero mode if a bulk field~- receives loop corrections 
via  the fields  $\phi_{M,k}$ and $\psi_{M,k}$.

\vspace{0.4cm}
\noindent
{\bf Case 3).} Continuous  Scherk-Schwarz supersymmetry breaking.
In this case, using the $SU(2)_R$ global symmetry, one can
impose continuous twists for the bulk fields
\begin{eqnarray}
\left(\begin{array}{l}\phi_M \\
\phi^{c\dagger}_M \end{array}\right)(x,y+2\pi R)
&=&e^{-2\pi i\omega\sigma_2}\left(\begin{array}{l}\phi_M \\
\phi^{c\dagger}_M \end{array}\right)(x,y), 
\end{eqnarray}
with $M=Q, U$. A similar transformation exists for
$(\lambda_1,\lambda_2)^T$ while $A_N(x,y)$, $N=\mu, 5$ and
$(\psi_M, \psi_M^c)^T$  do not acquire any twists under this
transformation. As a result, the fields $\phi_{M,k}$, $\phi_{M,k}^c$, 
will have mass $(k+\omega)/R$ while $\psi_{M,k}$, $(k\!\geq \!0)$ and
$\psi_{M,k}^c$ ($k\!\geq\! 1$) have masses $k/R$. For $\omega\!=\!0$
the fields $\psi_{M,k}$ and $\phi_{M,k}$ regain  equal masses at all levels.

\vspace{0.2cm}
\noindent
{\bf Case 4).} 
An additional orbifolding by $Z_2'$, so the orbifold is actually
$S_1/(Z_2\!\times\! Z_2')$. In this case the one-loop analysis is 
very close to  that of Case~2), since the $Z_2'$ action  has  similarities  to
$Z_{2,R}$ Scherk-Schwarz supersymmetry breaking.

With these considerations we  can present the one-loop  results to the 
mass of the scalar field $\phi_{H_u}$  induced by interaction
(\ref{interaction}).  For technical details see 
\cite{Ghilencea:2005hm,Ghilencea:2004sq}.
One obtains in all cases described  above the following
correction to the  mass of the scalar field $\phi_{H_u}$ (hereafter
denoted simply $\phi_{H}$)\footnote{In the case $\phi_{H_u}$ is a bulk field, 
the result refers to the one-loop  correction to the  mass of the zero
mode.}:
 
\begin{eqnarray}\label{mass}
-\, m_{\phi_H}^2 (q^2)\bigg\vert_B\!\!\!\! \!&=& \!\!\!\!
 \,  (2 f_{4, t})^2 N_c\!\!\! \sum_{k\geq 0,\, l\geq 0}
    \Big[\eta^{F_Q}_k \eta^{\phi_U}_l\Big]^2
   \!\! \int \frac{d^d p}{(2\pi)^d}
\frac{(-1)(p+q)^2\,\,\mu^{4-d}}{((p+q)^2+m_{\phi_Q^c,k}^2)(p^2+m_{\phi_U,l}^2)}
    \!+\! (Q\! \leftrightarrow\! U)
\nonumber\\[12pt]
- \, m_{\phi_H}^2 (q^2)\bigg\vert_F\!\!\!\!\!
 &= &\!\!\!\!   \,(2 f_{4, t})^2 N_c \!\sum_{k\geq 0,\, l\geq 0}
    \Big[\eta^{\psi_Q}_k \eta^{\psi_U}_l\Big]^2
    \!\int \frac{d^d p}{(2\pi)^d}
    \frac{2 \,p.(p+q)\,\,\mu^{4-d}}{((p+q)^2
+m_{\psi_Q,k}^2)(p^2+m_{\psi_U,l}^2)}\\
\nonumber
\end{eqnarray}
with $\mu$ the finite mass scale of the
 DR scheme. The Kaluza-Klein spectrum used above is 
\begin{eqnarray}
m_{\phi_Q,k}=\frac{1}{R}(k+c_1),&& \qquad 
m_{\phi_Q^c,k}=\frac{1}{R}(k+c_2),\qquad  k\geq 0
\nonumber\\[12pt]
m_{\psi_Q,k}=\frac{k}{R},\qquad\quad && 
\qquad m_{\psi_U,k}=\frac{k}{R}, \qquad \qquad \quad  k\geq 0
\end{eqnarray}
where the coefficients $c_{1,2}$ have values which depend on the
type of supersymmetry breaking: 
\begin{eqnarray}   
{\rm F-term\,\, breaking:}    && \!\!
c_1=1/2,\,\quad c_2=1, \,\,\,\,\,
\nonumber\\
{\rm Discrete\,\, Scherk-Schwarz:}   &&  \!\! c_1=1/2,\quad c_2=1/2, \,\,
\nonumber\\
{\rm Continuous \,\, Scherk-Schwarz:} &&  \!\! c_1=\omega,\,\,\,\,\,\quad
c_2=\omega,\,\quad\,\,
\nonumber\\
S_1/(Z_2\times Z_2'): &&  \!\! c_1=1/2,\quad c_2=1/2,
\end{eqnarray}

\noindent
In (\ref{mass}) one has the wavefunction coefficients 
$\eta_k^{\psi_M}={1}/{{\sqrt 2}^{\delta_{k,0}}}$, $M=Q,U$.
Also $\eta_k^{F_M}=\eta_l^{\phi_M}=1$ with the exception 
of the continuous Scherk-Schwarz case when
$\eta_k^{F_M}=\eta_l^{\phi_M}=1/\sqrt 2$ and when also the two 
sums in  the bosonic contribution  are over the whole set $\bZ$.
The result of the calculation of eq.(\ref{mass})  
for all cases  described is 

 \begin{eqnarray}\label{mass3}
-\, m_{\phi_H}^2 (q^2)
&=& \frac{(2 f_{4,t})^2}{2\, (4\pi R)^2} \, N_c\, \bigg\{\!
\int_0^1  dx \,
(2/\pi)\,\Big[ \cJ_2[0,0,c]-\cJ_2[c_1,c_2,c]\Big]
\nonumber\\[12pt]
&+&\,
\kappa_\epsilon (q^2 R^2)
\int_0^1  dx \Big[\,x \,(x-1)\,
\cJ_1[0,0,c]-(1-x)^2\,\cJ_1[c_1,c_2,c]\Big]\!\bigg\}
\end{eqnarray}
with $c=x (1-x) \,q^2 R^2$, $\kappa_\epsilon=(2\pi \mu R)^\epsilon$ and
\begin{eqnarray}\label{j12functions}
\cJ_j[ c_1,c_2,c]\!\! &\equiv &\!\!\!\!\!\!\sum_{k_1,k_2\in \bZ}\int_0^\infty
\!\!\!\frac{dt}{t^{j-\epsilon/2}}\,\, e^{-\pi\,
    t\,(c+a_1(k_1+c_1)^2+a_2(k_2+c_2)^2)}=
 \frac{\big(\!-\!\pi  c\,\big)^j}{j \sqrt{a_1 a_2}}\,
\bigg[\frac{2}{\epsilon}\bigg]\!+\cO(\epsilon^0),\,\,\,\,\, j\!=\!1,2.
\nonumber\\
\nonumber\\
&& a_1=(1-x), \qquad\, a_2= x, \qquad\, c=x (1-x) \,q^2 R^2.
\end{eqnarray}
Therefore the pole structure of $\cJ_j$, $j=1,2$ is independent of
the coefficients $c_1,c_2$  which distinguish between the four cases 
considered. The finite part $\cO(\epsilon^0)$ was also computed in 
\cite{Ghilencea:2005hm,Ghilencea:2004sq}.
If $q^2=0$ the second line in (\ref{mass3}) is
 absent, so $m^2_{\phi_H}(q^2=0)$ is
given by the first line alone. Further, the divergent part $c^2/\epsilon\sim 
q^4 R^2/\epsilon$ in each $\cJ_2$ cancels  in the difference
 $\cJ_2[0,0,c]-\cJ_2[c_1,c_2,c]$, but the divergence $c/\epsilon$
in both  $\cJ_1[0,0,c]$  and $\cJ_1[c_1,c_2,c]$
does not cancel in the second line in (\ref{mass3}).
One  then finds that for all models
\begin{eqnarray}\label{mass_final}
m_{\phi_H}^2(q^2)=m_{\phi_H}^2(0)-
\frac{(2 f_{4,t})^2}{2^8}
N_c\, (q^4 R^2)\, \bigg[\frac{1}{\epsilon}+\ln(2\pi R \mu)\bigg]
 +\frac{1}{R^2} \,\,\cO(q^2 R^2)
\end{eqnarray}
Let us address the significance of the 
above result. First,  $m^2_{\phi_H}(0)\sim f_{4,t}^2/R^2+\cO(\epsilon)$ 
where the
constant of proportionality depends on the exact details of
supersymmetry breaking (coefficients $c_{1,2}$).
Note that this constant can have a  negative sign 
 giving a negative one-loop $m^2_{\phi_H}(0)$. This can  induce electroweak
symmetry breaking  \cite{Delgado:2001si},
if the tree level mass of the scalar field is 
somehow arranged (by symmetry arguments) to vanish.

However,  the one-loop result of eq.(\ref{mass_final}) presents  a much broader
picture:  $m^2_{\phi_H}(q^2)$ is not finite, and the 
presence of the divergence $q^4 R^2/\epsilon$ 
which can only be ``seen'' at non-zero external momentum in the
associated two-point Green function, 
requires the addition to the Lagrangian of a counterterm
with four derivatives.  This can only be a brane-localised (N=1)
counterterm,  since 
 the interaction considered in eq.(\ref{interaction}) does not allow one to
write a bulk (N=2) counterterm that would necessarily involve
$H^c$ (which does not have a Yukawa coupling). Another explanation why
the counterterm is
brane-localised can be made using the mixed position-momentum two-point
Green functions in 5D \cite{Arkani-Hamed:2001mi}, which if
evaluated at $y=0$ give precisely
the result in eq.(\ref{mass}). Following these 
considerations  the counterterm has the structure (assuming $H_u$ is a
bulk field)
\begin{eqnarray}\label{hdo1}
\int \! d^4 x \,d y \!\!\int \! d^2 \theta \, d^2\overline \theta
\,\delta(y)\, \lambda_t^2 \, H_u^\dagger \Box  H_u
&\!\sim\!& f_{4,t}^2\!\!
\int\! d^4 x \,
 R^2 \!\!\! \sum_{n,p\geq 0} \phi_{H,n}^\dagger \Box^2  \phi_{H,p}
\nonumber\\
&\!\sim \!& f_{4,t}^2\!\!
\int \!\! d^4 x \,
 R^2 \, \phi_{H,0}^\dagger  \Box^2  \phi_{H,0}
\!+\!\cdots
\end{eqnarray}
with $[\lambda_t]=-3/2$.
However, if  $H_u$ is a brane field instead ($[H_u]=1$,
$[\lambda_t]=-1$), the counterterm reads 
\begin{eqnarray}\label{hdo2}
\int \! d^4 x \,d^2 \theta \, d^2\overline \theta \,\,
\lambda_t^2 \, H_u^\dagger \Box  H_u
\sim \! f_{4,t}^2\!
\int \!\! d^4 x \,
 R^2 \, \phi_{H}^\dagger  \Box^2  \phi_{H}
+...\\[-12pt]
\nonumber
\end{eqnarray}
We thus find that brane-localised higher derivative operators are generated 
by interaction (\ref{interaction}), as one-loop counterterms to the
mass of the (brane or zero mode of the bulk) scalar field. In the absence
of a detailed UV completion of the theory, one does not know the
overall (finite) coefficient in front of these operators. As a result 
the predictive
power of the models is significantly affected. Moreover, 
in the case of   higher order theories,  further complications may
arise, such as the generation of ghost fields, unitarity violation,
etc. \cite{Pais:1950za}. These observations underline the importance of 
the study of such operators and the need for a UV completion of such theories.

The generation of higher
derivative counterterms is  solely due to compactification.
To understand this note that in eq.(\ref{mass}) we summed
over the whole  Kaluza-Klein towers of modes, associated with the compact
dimension. This was done to respect the discrete shift
symmetry $k_i\rightarrow k_i+1$  present in eqs.(\ref{mass3}),
(\ref{j12functions}), which are just  a re-writing of the initial
 sums  in (\ref{mass}). But to illustrate the origin of the one-loop higher
derivative counterterms, it is instructive to examine what happens when
the two sums in  (\ref{mass}) or 
 (\ref{j12functions})  were truncated each to a finite but otherwise 
arbitrary Kaluza-Klein level (while respecting the $N=2$ multiplet
structure of the modes). The 
framework would then be that of a 4D theory with a finite number of
Kaluza-Klein states, i.e. a renormalisable one. 
In that case, the two sums in  $\cJ_{j}$  ($j=1,2$) truncated to say
$s_1$ and $s_2$,  would diverge as $\cJ_j\sim s_1 s_2/\epsilon$ 
\cite{Ghilencea:2004sq}. This is to be
compared with the divergence $c^j/\epsilon$, $j=1,2$ of ``untruncated'' 
$\cJ_j$, eq.(\ref{j12functions}).  Using this, one would obtain
from eq.(\ref{mass3})  divergences of type  
$m^2_{\phi_H}(q^2)\sim m^2_{\phi_H}(0)+s_1 s_2 \, q^2/\epsilon+...$
but no $q^4 R^2/\epsilon$ terms. The terms
$s_1 s_2 \,q^2/\epsilon$ would  then   account for 
wavefunction renormalisation only; in addition $m^2_{\phi_H}(0)$  
would  have the usual 4D quadratic divergence rather than being
one-loop finite 
(as it was  when summing over the whole towers).
 Therefore higher derivative operators, related to the presence 
of $q^4 R^2/\epsilon$ only emerge when summing over the whole 
Kaluza-Klein towers, so they are indeed the result of compactification. 
Their presence is due to  the   same reason
which enforced a one-loop finite~$m^2_{\phi_H}(0)$.

Let us return to the analysis of eq.(\ref{mass_final}). After adding 
the counterterm
in the action one ends with a similar equation, but the pole 
$1/\epsilon$ is replaced by an unknown, finite coefficient (hereafter
denoted $\xi$). In this equation,  if $R$ is somehow fixed
to a large value (inverse TeV scale) in order  to have a small 
mass for the scalar field (without large fine tuning), 
 the second term in this equation, $\xi q^4 R^2$, becomes more
important. Given the unknown value of $\xi$ 
this affects significantly  the predictive power 
of the models. Conversely, if $R$ is very small ($q^2$ fixed),
the role of higher derivative operators is suppressed, but
the first term $\sim 1/R^2$
re-introduces the quadratic mass scale (hierarchy) problem
at one-loop,  familiar from the Standard Model. While this is
the general picture, a detailed analysis should also consider the $\cO(q^2
R^2)$  terms   in (\ref{j12functions}). 
Finally, our  calculation can be used to  re-address previously
mentioned studies of the radiative electroweak symmetry breaking
 induced by towers
of Kaluza-Klein modes, which ignored the effect of the higher
derivative operators. Our results eqs.(\ref{mass3})-(\ref{mass_final})
also provide  the  running of the  scalar mass and its UV behaviour
 under the UV
scaling of the   momenta, $q^2\ra \rho\, q^2$, $\rho\gg 1$.

We can now  address the role (the initial 5D N=1) supersymmetry
 plays in this calculation. According to eq.(\ref{mass3}), the
 divergences $q^4 R^2/\epsilon$ present in each $\cJ_2$ 
cancel in the difference in the first line of eq.(\ref{mass3})
which contributes to $m^2_{\phi_H}(0)$. Thus initial supersymmetry and
the summing over the whole Kaluza-Klein towers lead to one-loop
 finite $m^2_{\phi_H}(0)$. Note however that
 an {\it identical}  divergence $q^4 R^2/\epsilon$
 originating from the second line of eq.(\ref{mass3}) and due to 
$\cJ_1$ functions does not cancel, leading to the need for higher
 derivative counterterms. Thus  initial
 supersymmetry did not protect
against the generation at the one-loop level of such counterterms,
 and this  is true regardless of the values of $c_{1,2}$ i.e.
 of the way supersymmetry was broken
(discrete/continuous Scherk-Schwarz mechanism,  F-term breaking or
$S_1/(Z_2\times Z_2')$). Ultimately, such conclusion may not be
too surprising if we recall that initial theory, although supersymmetric
is nevertheless non-renormalisable.
While (initial) supersymmetry ensured a cutoff independent $m^2(0)\sim 1/R^2$, 
higher derivative operators are present, similarly to  any non-supersymmetric
effective  theory 
in which the UV cutoff of the theory is replacing the scale $1/R^2$.

These observations are  important since one would naively expect that
supersymmetry would ensure that after compactification  higher derivative
operators and their consequences on the mass of the scalar field 
would be  somewhat under control. But the unknown coefficient $\xi$
 of these operators  also introduces a new
parameter i.e. scale in the theory where such operators become
 important and which is
relevant for the loop corrected mass of the scalar field.
One has to fix this scale and also $1/R^2$  by some dynamical
mechanism, to provide a  solution to the hierarchy  problem 
in the context of higher dimensional  theories.

The analysis has so far  considered a localised
superpotential interaction $\lambda_t\, Q\, U\, H_u$ where the fields
$Q, U$ were bulk fields, while $H_u$ was either a bulk  or a brane field.
Other possibilities for the character brane/bulk of these fields
may be  allowed. Then
 dimensional arguments allow us to estimate the order $n$ in
perturbation theory when the (localised) higher derivative counterterm 
$(\lambda_t^2)^n H_u^\dagger \Box H_u$ is generated.
If the  interaction has two genuine brane
fields and one bulk  field, $[\lambda_t]=-1/2$. Then,  if $H_u$ is a 
 brane field, one has $n=2$ so the local counterterm is generated  
at the two-loop level. 
If $H_u$ is the (only) bulk field, dimensional arguments give
that  $n=3$, thus such counterterm may arise at three-loop only.
Similar considerations can be made for the bulk (gauge) 
interactions  using that the gauge coupling also has mass 
dimension $-1/2$. These observations can be used when building 
higher dimensional models, to avoid such counterterms at  
small number of loops.

\section{Conclusions.}

In  5D N=1 supersymmetric models compactified on $S_1/Z_2$
or $S_1/(Z_2\times Z_2')$ we  discussed the loop corrections that 
a (brane-localised) superpotential induces to the mass of the 
(brane or zero mode of the bulk) scalar field. Such interaction is 
very common in most higher dimensional extensions of the SM or MSSM
models and the scalar field is usually regarded as the Higgs field candidate.
The analysis investigated the link between the nature of supersymmetry
breaking on these orbifolds and the  emergence  of higher derivative 
counterterms to the scalar field mass. Gauge interactions can also
induce such  higher derivative operators at one-loop (for example in
 6D  \cite{dmghml}), 
but for the 5D case  this arises beyond the one-loop order.

It was found that (brane-localised) higher derivative counterterms to the
mass  of the Higgs field 
are generated at the one loop level. As a result the mass of the 
scalar field depends on the unknown coefficient $\xi$ that such operators
come with in the action. The mass of the scalar field behaves like
$m^2_{\phi_H}(q^2)=m^2_{\phi_H}(0)+\xi\, q^4 R^2+ 1/R^2\, \cO(q^2 R^2)$ 
with $m^2_{\phi_H}(0)\sim f_{4,t}^2/R^2$. Note that a somewhat similar
structure can emerge even in the SM with additional,  higher
derivative operators, but with $1/R$ replaced by the UV cutoff of the model.
With $q^2$ fixed, if $R$ is small one would expect
 the higher derivative operators have very small
effects, and uncertainties induced by the coefficient $\xi$ are suppressed.
However, in this case   $1/R^2$ is
very large and  one restores the  usual mass hierarchy problem of the SM.
Alternatively, if $R$ is large (TeV region) one may have a small
$m^2_{\phi_H}(q^2)$ (at $q^2\sim m^2$) and thus an electroweak scale
mass for the scalar field without large fine tuning. However, in that
case $\xi \,q^4 R^2$ terms become more important and re-introduce 
uncertainty in the prediction, due to $\xi$ or
equivalently,  unknown physics {\it above}  the compactification scale.
The value of $\xi$ and $1/R^2$  must  be fixed by a
dynamical  mechanism in order to solve the hierarchy problem in 
higher dimensional supersymmetric models compactified on $S_1/Z_2$ or
$S_1/(Z_2\times Z_2')$.

We investigated the relation between supersymmetry breaking and the
presence of higher derivative operators. 
It turned out that such operators are present regardless
of the supersymmetry breaking mechanism  considered (F-term
breaking, discrete or continuous Scherk-Schwarz mechanisms, additional
$Z_2'$ orbifolding).
Therefore (initial) supersymmetry does not protect against the presence of such
operators which were shown to be ultimately due to  a divergence 
identical to  that cancelled in $m_{\phi_H}(0)$ by (initial) supersymmetry.

The presence of the higher derivative  operators is directly
related to the number of bulk fields involved in the interaction. In our case 
the  counterterms were ultimately generated because of the (two) 
corresponding Kaluza-Klein sums  acting on the loop integral. 
They emerged from  a ``mixing'' effect  between one infinite 
sum  and a winding zero-mode (on the lattice dual to that of Kaluza-Klein
modes of the second sum). For this reason such operators  can be 
considered  of  non-perturbative, non-local origin and are a generic
presence in models with extra dimensions.

Gauge theories with higher derivative operators in the action were not in
general the most popular in the past, due to the fact that such operators may
bring  in a host of complications,  related to the presence of
additional ghost fields,  unitarity violation or non-locality effects.
Nevertheless, the emergence of such operators from compactification as  
counterterms to the masses (and/or the couplings) of the models, 
shows that these operators are very important at the quantum level.
To conclude,  addressing the role that higher derivative operators 
play in orbifold models,  for  the hierarchy problem in particular,
is an interesting study  given that such operators are
a   common presence in  higher dimensional models.

\section*{Acknowledgements:} 
\noindent
The work of D.G. was supported by a research grant from PPARC, United Kingdom.

\end{document}